\documentclass[12pt,a4paper,reqno]{amsart}

\usepackage{amsthm,a4}
\usepackage{esint}
\usepackage{hyperref}
\usepackage[british]{babel}

\newtheorem{theorem}{Theorem}
\newtheorem{question}[theorem]{Question}

\newcommand{\dd}{\mathrm{d}}
\newcommand{\ee}{\mathrm{e}}
\DeclareMathOperator{\Res}{Res}
\DeclareMathOperator{\Pf}{Pf}
\DeclareMathOperator{\Tr}{Tr}
\DeclareMathOperator{\diag}{diag}
\newcommand{\bs}[1]{\ensuremath{\boldsymbol{#1}}}

\allowdisplaybreaks[4]

\begin{document}

\title[Linear relations supplementing BKP]{Linear relations supplementing BKP for the Kontsevich matrix model
  with even potential}

\date{}
\author{Raimar Wulkenhaar}
\address{Mathematisches Institut, Universit\"at M\"unster,
  Einsteinstr.\ 62, 48149 M\"unster, Germany.}
\email{raimar@math.uni-muenster.de}

\begin{abstract}
  With G. Borot we have previously proved that moments of the
  Kontsevich matrix model with arbitrary potential satisfy quadratic
  BKP relations. In this note we show that, for even potential, these
  moments also satisfy simpler linear relations, which are organised by
  elementary Schur Q-functions.
\end{abstract}


\maketitle

\section{Preliminaries}

Let $\mathcal{H}_N$ be the space of Hermitian $N \times N$ matrices,
equipped with Lebesgue measure
$
\dd H= \prod_{i=1}^N \dd H_{ii}
  \prod_{1 \leq i < j \leq N} \dd\mathrm{Re}(H_{ij})\, \dd\mathrm{Im}(H_{ij})$.
For a positive matrix $\Lambda = \diag (\lambda_1,\ldots,\lambda_N)$
we introduce the Gau{\ss}ian probability measure on $\mathcal{H}_N$
\begin{equation}
\label{theGauss}
\dd \mathbb{P}_\Lambda(H) = \frac{\sqrt{ \prod_{1 \leq i,j
      \leq N} (\lambda_i + \lambda_j)  }}{
  2^{\frac{N}{2}} (2\pi)^{\frac{N^2}{2}}}\,
\,\ee^{-\frac{1}{2}{\rm Tr}(\Lambda H^2)}\,\dd H.
\end{equation}
For a continuous function $V_0$ on
$\mathbb{R}$ such that the measure
$\ee^{- \frac{1}{2}\lambda_i x^2 + V_0(x)}$ has finite moments on
$\mathbb{R}$ for every $1\leq i\leq N$, and a family
$\bs{t} = (t_{2k + 1})_{k \geq 0}$ of formal parameters, we define
\begin{align}
V_{\bs{t}}(x) = V_0(x) +  \sum_{k \geq 0} t_{2k + 1}\,x^{2k + 1}.
\end{align}
Then measure and partition function of the Kontsevich matrix
model with arbitrary potential are defined by
\begin{equation}
  \label{Zdef}
  \dd\mu_{\Lambda,\bs{t}}(H) =  
  \ee^{\Tr (V_{\bs{t}}(H))} \dd\mathbb{P}_\Lambda(H),\qquad
  Z_\Lambda(\bs{t}) = \int_{\mathcal{H}_N}   \dd\mu_{\Lambda,\bs{t}}(H) .
\end{equation}
In \cite{Borot:2023thu} the following results were established
(assuming that $N$ is even):
\begin{theorem}[{\cite{Borot:2023thu}}]
\label{thm:BW24}
\textup{(A)}\quad The partition function $Z_\Lambda(\bs{t})$ is a
$\tau$-function of the BKP integrable hierarchy \cite{Date:1981qz},
i.e.\ it satisfies the B-type Hirota bilinear relation
\begin{equation}
\label{biline}
Z_\Lambda(\bs{t})Z_\Lambda(\tilde{\bs{t}}) = \Res\displaylimits_{z = 0}\,
\frac{\dd z}{z}\, \ee^{\sum_{k \geq 0} z^{2k + 1}(t_{2k + 1}
  - \tilde{t}_{2k + 1})} Z_\Lambda\big(\bs{t}
- 2[z^{-1}]\big)\,Z_\Lambda\big(\tilde{\bs{t}} + 2[z^{-1}]\big)
\end{equation}
identically in $\bs{t},\tilde{\bs{t}}$,
where $[z^{-1}] = \big(\frac{1}{z},\frac{1}{3z^3},\frac{1}{5z^5},\ldots\big)$.   

\medskip

\noindent
\textup{(B)}\quad The partition function evaluates to
\begin{align}
\label{Ztint}
Z_\Lambda(\bs{t}) 
& = \frac{C(\Lambda)}{N!} \int_{\mathbb{R}^N}
\Pf\displaylimits_{1\leq i,j\leq N} \Big(\frac{1}{2}\cdot
\frac{x_j-x_i}{x_j+x_i}\Big)
\det\displaylimits_{1 \leq i,j \leq N} \big(\ee^{-\frac{1}{2}\lambda_ix_j^2}\big)\,
\prod_{i = 1}^N \ee^{V_{\bs{t}}(x_i)} \dd x_i
\\
&= C(\Lambda) \Pf\displaylimits_{1\leq i,j\leq N}
\bigg(\fint_{\mathbb{R}^2} \frac{1}{2} \cdot \frac{x - y}{x + y}\,
\ee^{-\frac{1}{2}\lambda_i x^2 - \frac{1}{2}\lambda_j y^2
  + V_{\bs{t}}(x) + V_{\bs{t}}(y)} \,\dd x \dd y\bigg),
\nonumber
\end{align}
where $C(\Lambda):=\frac{(-1)^{\frac{N(N - 1)}{2}}
  \sqrt{ \prod_{1 \leq i,j
        \leq N} (\lambda_i + \lambda_j)
}}{  (2\pi)^{\frac{N}{2}}\,  \Delta(\boldsymbol{\lambda})}$.
Here $\Delta(\boldsymbol{\lambda})
= \prod_{1 \leq i < j \leq N}  (\lambda_j - \lambda_i)$ is the 
Vandermonde determinant 
and $\fint$ the Cauchy principal value integral.
\end{theorem}

The Hirota relation (\ref{biline}) has several equivalent expressions;
most of them involve the elementary Schur-Q functions (see e.g.\
\cite[Appendix D.5]{harnad_balogh_2021})
\begin{align}
q_j(\bs{t}):= [x^j]\exp\big(2\sum_{k=0}^\infty t_{2k+1} x^{2k+1} \big).
\label{def:q}
\end{align}
Setting $\bs{t}\mapsto \bs{t}+\bs{s}$ and 
$\tilde{\bs{t}}\mapsto \bs{t}-\bs{s}$ and using
(\ref{def:q}) one evaluates (\ref{biline}) to
\begin{align}
0 &=\sum_{\substack{k,l,m,n,r,s=0\\k+l\neq 0}}^\infty(-1)^{k+n}
q_{k+l}(\bs{s})
\label{BKP-ZN-q}
\\
& \times
\int_{\mathcal{H}_N} \!\!\! \dd\mu_{\Lambda,\bs{0}}(H)\,
q_k(\tilde{\bs{H}})
q_m(\tfrac{1}{2}\bs{sH})
q_r(\tfrac{1}{2}\bs{tH})
\int_{\mathcal{H}_N} \!\!\! \dd\mu_{\Lambda,\bs{0}}(H)\,
q_l(\tilde{\bs{H}})
q_n(\tfrac{1}{2}\bs{sH})
q_s(\tfrac{1}{2}\bs{tH})
\nonumber
\end{align}
identically in $\bs{s},\bs{t}$,
where we defined $\tilde{\bs{H}}
:=(\Tr(H),\frac{1}{3} \Tr(H^3),\frac{1}{5} \Tr(H^5),...)$ and
$\bs{sH}:=(s_1\Tr(H), s_3 \Tr(H^3), s_5 \Tr(H^5),...)$.

The BKP hierarchy is conveniently formulated in terms of neutral
fermionic operators $\{\phi_k\}_{k\in \mathbb{Z}}$ which satisfy the
anticommutation relation $\{\phi_k,\phi_l\}=(-1)^k \delta_{k+l,0}$.
There is a vacuum vector $|0\rangle$ which satisfies
$\langle 0|0\rangle=1$, $\phi_{-k}|0\rangle=0$ and
$\langle 0|\phi_{k}=0$ for $k>0$ as well as
$\langle 0|\phi_{0}|0\rangle=0$. We also introduce the generating
series $\phi(x)=\sum_{k\in \mathbb{Z}} \phi_k x^k$.  Vacuum
expectations of products of $\phi(x_i)$ are understood in a radial
ordering. If all $|x_i|$ are pairwise distinct, then
\begin{equation}
\label{radialordering}
  \begin{split}
  \langle 0|\phi(x_1)\cdots   \phi(x_{N})|0\rangle
&:=(-1)^{\mathrm{sign}(\pi)}
\langle 0|\phi(x_{\pi(1)})\cdots   \phi(x_{\pi(n)})|0\rangle
\\
\text{if} \qquad & |x_{\pi(1)}|>|x_{\pi(2)}|>\dots >|x_{\pi(N)}|.
\end{split}
\end{equation}
One finds
\begin{equation}
 \label{pair01} \langle 0|\phi(x_1)\phi(x_2)|0\rangle
  =\frac{1}{2} \,\frac{x_1-x_2}{x_1+x_2},
\end{equation}
understood as convergent power series in
$\frac{x_1}{x_2}$ for $|x_1|<|x_2|$ and as convergent power series in
$\frac{x_2}{x_1}$ for $|x_2|<|x_1|$. The following is known as Wick's theorem:
For pairwise different $|x_i|$ one has for $N$ even
\begin{equation}
\label{Wick}
\begin{split}
  \langle 0|\phi(x_1)\phi(x_2)\cdots
  \phi(x_{N})|0\rangle
  & =\mathop{{\rm Pf}}_{1\leq k,l \leq N}
  \Big(\langle 0|\phi(x_k)\phi(x_l)|0\rangle
  \Big) \\
  &= \frac{1}{2^{\frac{N}{2}}}
  \mathop{{\rm Pf}}_{1\leq k,l \leq N}
  \Big(\frac{x_k-x_l}{x_k+x_l}\Big)
= \frac{1}{2^{\frac{N}{2}}}\prod_{1\leq k<l\leq N}
\frac{x_k-x_l}{x_k+x_l},
\end{split}
\end{equation}
and for $N$ odd
$\langle 0|\phi(x_1)\phi(x_2)\cdots \phi(x_{N})|0\rangle=0$. See e.g.\
\cite[sec.~7.3]{harnad_balogh_2021} for details.

\section{Linear relations for even potential $V_0(x)=V_0(-x)$}

We prove that for any \emph{even} potential $V_0(x)=V_0(-x)$
we have, in addition to the bilinear relations (\ref{BKP-ZN-q}),
also a family of \emph{linear relations}:
\begin{theorem}
Let  $[z^{-1}]=(\frac{1}{z},\frac{1}{3z^3},\frac{1}{5z^5},...)$
and 
$\tilde{\bs{H}}
=(\Tr(H),\frac{1}{3} \Tr(H^3),\frac{1}{5} \Tr(H^5),...)$
for $H\in \mathcal{H}_N$.
Let $V_0$ be as above. The measure
$\dd\mu_{\Lambda,\bs{t}}(H)$ and the partition function
$Z_\Lambda(\bs{t})$ obey the following relations:

\medskip

\noindent
\textup{(A)}\quad For any $k\in \mathbb{Z}_{>0}$ one has
\begin{align}
  \label{int-qm}
(-1)^k 
  \int_{\mathcal{H}_N}
  \!\!\!\dd\mu_{\Lambda,\bs{t}}(H)
  q_{k}(\tilde{\bs{H}})
&\equiv  
\Res\displaylimits_{z=0} \Big( Z_\Lambda(\bs{t}-2[z^{-1}])
z^{k-1} \dd z\Big)
\\
  &= \sum_{1\leq r<s\leq N} \!\!\!
Z_{\Lambda_{r,s}}(\bs{t})
\frac{  \sqrt{\lambda_r\lambda_s}}{\pi}
\frac{\lambda_s+\lambda_r}{\lambda_s-\lambda_r}
\prod_{\substack{i=1\\ i\neq r,s}}^N
\frac{(\lambda_r+\lambda_i)(\lambda_s+\lambda_i)}{
(\lambda_r-\lambda_i)(\lambda_s-\lambda_i)}
  \nonumber
  \\*[-1ex]
    &\times 
    \int \dd x \dd y\,(-1)^k(x^k-y^k)
\, \ee^{-\frac{1}{2} \lambda_s x^2-\frac{1}{2} \lambda_r y^2}
\ee^{V_{\bs{t}}(x)+V_{\bs{t}}(y)},
  \nonumber
\end{align}
where
$\Lambda_{r,s}=\diag (\lambda_1,\stackrel{r,s}{\check{\ldots}},\lambda_N)
\in \mathcal{H}_{N-2}$ is the submatrix of $\Lambda$
with $r^{th},s^{th}$ rows and
columns deleted, for $1\leq r<s\leq N$.

\medskip

\noindent
\textup{(B)}\quad For even $V_0(x)=V_0(-x)$ one has,
as formal power series identically in $\bs{t}=(t_1,t_3,t_5,...)$,
\begin{align}
0&=\Res\displaylimits_{z=0} \Big( Z_\Lambda(\bs{t}-2[z^{-1}])
  \Big(\ee^{2\sum_{j=0}^\infty t_{2j+1} z^{2j+1}}-1\Big)
  \frac{\dd z}{z}\Big)
  \nonumber
  \\
&= \sum_{k=1}^\infty \sum_{m=0}^\infty (-1)^k q_k(\bs{t})
\int_{\mathcal{H}_N} \!\!\! \dd\mu_{\Lambda,\bs{0}}(H)\,
q_k(\tilde{\bs{H}})
q_m(\tfrac{1}{2}\bs{tH}) .
\label{Zt-linear}
\end{align}

\noindent
\textup{(C)}\quad For even $V_0(x)=V_0(-x)$ and any odd $k,m\geq 1$ one has
  \begin{align}
\label{int-even}
    \int_{\mathcal{H}_N} \!\!\!\dd\mu_{\Lambda,\bs{0}}(H)
q_{k+m}(\tilde{\bs{H}}) 
&=    \int_{\mathcal{H}_N} \!\!\!\dd\mu_{\Lambda,\bs{0}}(H)
\Tr(H^m)q_{k}(\tilde{\bs{H}}) .
\end{align}
\begin{proof}
  (A)\quad We have
$\dd\mu_{\Lambda,\bs{t}-2[z^{-1}]}(H)=
\dd\mu_{\Lambda,\bs{t}}(H)\exp\big(2 \sum_{j=0}^\infty
(\frac{-1}{z})^{2j+1} \frac{\mathrm {Tr}(H^{2j+1})}{2j+1}
\big)$.
Expressing the exponential by
(\ref{def:q}) for 
$x=-\frac{1}{z}$ and
$\bs{t}=\tilde{\bs{H}}$, one  
finds the first line of (\ref{int-qm}),
$    \Res\displaylimits_{z=0} \big( Z_\Lambda(\bs{t}-2[z^{-1}])
z^{k-1} \dd z\big)
=(-1)^k
\int_{\mathcal{H}_N} \dd\mu_{\Lambda,\bs{t}}(H) q_{k}(\tilde{\bs{H}})$.
We rewrite  $Z_\Lambda$ given in the
first line of (\ref{Ztint}) using the Wick products (\ref{Wick})
of fermionic operators. The shift by $-2[z^{-1}]$ can be implemented by  
operators
$\gamma(\bs{t})=\ee^{\sum_{j=0}^\infty t_{2j+1}J_{2j+1}}$
with $[J_{2k+1},J_{2l+1}]=0$ and $[J_{2k+1},\phi(x)]=x^{2k+1}\phi(x)$
as well as  $J_{2k+1}|0\rangle=0$:
\begin{align}
  Z_\Lambda(\bs{t}-2[z^{-1}])=\frac{C(\Lambda)}{N!}
  \int_{\mathbb{R}^N} &
  \langle 0|\gamma(-2[z^{-1}]) \phi(x_1)\cdots \phi(x_N)|0\rangle
  \nonumber
  \\*[-2ex]
  & \times \det_{1\leq i,j\leq N} (\ee^{-\frac{1}{2} \lambda_ix_j^2})
  \prod_{n=1}^N \ee^{V_{\bs{t}}(x_n)} \dd x_n\;.
\label{ZN-det}
\end{align}
Indeed, commuting $\gamma(-2[z^{-1}])$ to the right 
in front of the vacuum $|0\rangle$ 
gives a factor 
$\prod_{n=1}^N\ee^{-2\sum_{j=0}^\infty \frac{1}{(2j+1)z^{2j+1}}x_n^{2j+1}}$.
Lemma 7.3.9 in \cite{harnad_balogh_2021} states
$\langle 0|\gamma(-2[z^{-1}])=2\langle 0|\phi_0\phi(z)$. This allows us to
evaluate the residue for $k\geq 1$
\begin{align*}
\Res\displaylimits _{z=0} &
\langle 0|\gamma(-2[z^{-1}]) \phi(x_1)\cdots \phi(x_N)|0\rangle
z^{k-1}\dd z
\\
&= 2
\langle 0|\phi_0 \phi_{-k} \phi(x_1)\cdots \phi(x_N)|0\rangle
\\
&=\sum_{1\leq p<q\leq N} (-1)^{p+q} (-1)^k(x_q^k-x_p^k) 
\langle 0|\phi(x_1)\stackrel{p,q}{\check{\cdots}} \phi(x_N)|0\rangle.
\end{align*}
The final line follows from Wick's theorem (\ref{Wick}) and the
recursive property of a Pfaffian, together with
$\langle 0|\phi_0\phi_{-k}|0\rangle=0$
and $\langle 0|\phi_{-k} \phi(x_p)|0\rangle=(-1)^k x_p^k$
for $k>0$ as well as
$\langle 0|\phi_0 \phi(x_q)|0\rangle=\frac{1}{2}$.
The remainder is an
expectation value/Pfaffian of size $N-2$ (with $\phi(x_p)$ and
$\phi(x_q)$ left out).

We expand the determinant in (\ref{ZN-det}) with respect to
rows $q$ (first) and $p$:
\begin{align}
  \det_{1\leq i,j\leq N} (\ee^{-\frac{1}{2} \lambda_ix_j^2})
= \sum_{1\leq r<s \leq N}
(-1)^{p+q+r+s}&
\big(\ee^{-\frac{1}{2} \lambda_s x_q^2-\frac{1}{2} \lambda_r x_p^2}
-\ee^{-\frac{1}{2} \lambda_r x_q^2-\frac{1}{2} \lambda_s x_p^2}  \big)
\nonumber
\\*[-2.5ex]
& \times 
\det_{\substack{1\leq i,j\leq N \\ i\neq r,s ,~j\neq p,q}}
(\ee^{-\frac{1}{2} \lambda_ix_j^2}).
\label{det-rs}
\end{align}
Finally, we isolate the dependence on $\lambda_r,\lambda_s$ in the
prefactor $C(\Lambda)$ in Thm.~\ref{thm:BW24}:
\begin{align*}
C(\Lambda) &=(-1)^{r+s}C(\Lambda_{r,s})
\frac{\sqrt{\lambda_r\lambda_s}}{\pi}
\frac{\lambda_s+\lambda_r}{\lambda_s-\lambda_r}
\prod_{\substack{i=1\\ i\neq r,s}}^N
\frac{(\lambda_r+\lambda_i)(\lambda_s+\lambda_i)}{
  (\lambda_r-\lambda_i)(\lambda_s-\lambda_i)}.
\end{align*}
We put everything together, where we change the order of summations from
$\sum_{1\leq p<q\leq N} \sum_{1\leq r<s\leq N}$ to
$\sum_{1\leq r<s\leq N} \sum_{1\leq p<q\leq N}$.
After renaming $x_q\mapsto x$, $x_p\mapsto y$ and shifting  
the indices of $x_i$ with $i>p$, every of
the $\frac{N(N-1)}{2}$ pairs $p<q$ gives the same contribution.
After a final exchange $x\leftrightarrow y$ in the difference
(\ref{det-rs}) we confirm the result (\ref{int-qm}).

\medskip

\noindent
(B)\quad We multiply (\ref{int-qm})
by $q_k(\bs{t})$ and sum over $k\geq 1$. Using the 
identity (\ref{def:q}) we can partly resum to exponentials, e.g.\ 
$\sum_{k=1}^\infty (-1)^kq_k(\bs{t}) (x^k-y^k)=
\ee^{-2\sum_{j=0}^\infty t_{2j+1}x^{2j+1}}-\ee^{-2\sum_{j=0}^\infty t_{2j+1}y^{2j+1}}$,
and obtain
\begin{align}
  &  \sum_{k=1}^\infty 
  \int_{\mathcal{H}_N}
  \!\!\!\dd\mu_{\Lambda,\bs{t}}(H)
(-1)^k q_k(\bs{t}) q_{k}(\tilde{\bs{H}})
\label{Zt-linear-B1}
\\
&\equiv \Res\displaylimits_{z=0} \Big( Z_\Lambda(\bs{t}-2[z^{-1}])
  \Big(\ee^{2\sum_{j=0}^\infty t_{2j+1} z^{2j+1}}-1\Big)
  \frac{\dd z}{z}\Big)
  \nonumber
  \\
  &= \sum_{1\leq r<s\leq N}
Z_{\Lambda_{r,s}}(\bs{t})
\frac{  \sqrt{\lambda_r\lambda_s}}{\pi}
\frac{\lambda_s+\lambda_r}{\lambda_s-\lambda_r}
\prod_{\substack{i=1\\ i\neq r,s}}^N
\frac{(\lambda_r+\lambda_i)(\lambda_s+\lambda_i)
}{(\lambda_r-\lambda_i)(\lambda_s-\lambda_i)}
  \nonumber
  \\*[-1ex]
    &\times 2
    \int \dd x \dd y\,
    \, \ee^{-\frac{1}{2} \lambda_s x^2-\frac{1}{2} \lambda_r y^2
      +V_{\bs{0}}(x)+V_{\bs{0}}(y)}
\sinh \Big(\sum_{j=0}^\infty t_{2j+1}(y^{2j+1}-x^{2j+1})\Big).
  \nonumber
\end{align}
So far this is true for any potential $V_0=V_{\bs{0}}$
that admits finite moments.
For even $V_0(x)=V_0(-x)$ an odd
function under the transformation $T(x,y)= (-x,-y)$ is integrated over 
the $T$-symmetric domain $\mathbb{R}^2$, so that
(\ref{Zt-linear-B1}) vanishes identically. Expanding 
$\dd\mu_{\Lambda,\bs{t}}(H)$ via (\ref{def:q}) in $\bs{t}$
we confirm (\ref{Zt-linear}).

\medskip

\noindent
(C)\quad For even $V_0(x)=V_0(-x)$ and odd natural numbers $k,m$ we compare:
\begin{enumerate}
\item[(i)] 
  On one hand we take $k\mapsto k+m$ even in
  (\ref{int-qm}) at $\bs{t}=\bs{0}$.

\item[(ii)]  On the other hand we differentiate 
(\ref{int-qm}) with respect to $t_m$ and
set $\bs{t}=\bs{0}$. Since
$\frac{\partial Z_{\Lambda_{r,s}}(\bs{t})}{\partial t_m}
\big|_{\bs{t}=\bs{0}}=0$ for even potential $V_0$ and odd $m$, only the
case where the $t_m$-derivatives acts on the $(x,y)$-integral in the
last line of (\ref{int-qm}) contributes. The $t_m$-derivative simply
yields a factor $(x^m+y^m)$ under the integral.
At $\bs{t}=\bs{0}$, because of $V_0(x)=V_0(-x)$, we can set
$(x^{k}-y^{k})(x^m+y^m)\mapsto 
(x^{k+m}-y^{k+m})$ under the integral (the odd functions $x^my^k$
and $x^ky^m$ integrate to zero), which coincides with the 
evaluation in case (i).
\end{enumerate}
This finishes the proof of (\ref{int-even}).
\end{proof}
\end{theorem}

\section{Relations up to order 8 and open question}

We list the BKP and linear relations up to order 8, expressed in
terms of moments 
\begin{align}
  M_{n_1^{r_1},n_2^{r_2},\dots,n_p^{r_p}}
  = \frac{1}{Z_\Lambda( \bs{0})}
   \int_{\mathcal{H}_N} \dd\mu_{\Lambda,\bs{0}}\,\Big(
(\Tr(H^{n_1}))^{r_1}
(\Tr(H^{n_2}))^{r_2}\cdots
(\Tr(H^{n_p}))^{r_p}\Big),
\label{moments}
\end{align}
where $Z_\Lambda( \bs{0})=\int_{\mathcal{H}_N} \dd\mu_{\Lambda,\bs{0}}$.
Here, $(n_1,...,n_p)$ is a weakly decreasing tuple of
odd natural numbers, and for the exponents
$r_i\geq 1$ we agree to omit $r_i=1$.
For \emph{even potential} $V_0(x)=V_0(-x)$ we
have $M_{2\ell_1 + 1,\ldots,2\ell_n + 1} = 0$
for $n$ odd.

The  BKP Hirota relation (\ref{BKP-ZN-q})
reads for even potential $V_0(x)=V_0(-x)$ at $\bs{t}=\bs{0}$
up to order 8: 
\begin{subequations}
\label{BKP-even}
  \begin{align}
    0 & =-\frac{16}{2025}\big(45 s_5s_1-45 s_3^2-15 s_3s_1^3+s_1^6)
    \nonumber
    \\
    & \times  \Big\{ \big(9 M_{5,1} - 5M_{3^2} - 5 M_{3,1^3}  + M_{1^6}  
    \big)
  -15 M_{1^2} \big(M_{3,1} - M_{1^4} \big)\Big\}
\\
& - \frac{8}{99225}
\big(630 s_7s_1-630 s_5s_3-105 s_5s_1^3
-315 s_3^2s_1^2+21 s_3s_1^5+s_1^8\big)
    \nonumber
\\
&\times \Big\{ \big(90 M_{7,1} - 42M_{5,3} - 21 M_{5,1^3}
- 35 M_{3^2,1^2} + 7M_{3,1^5} +  M_{1^8}  \big)
    \nonumber
\\
&
\qquad-7 M_{1^2} \big(9 M_{5,1} + 5 M_{3^2}- 10M_{3,1^3} - 4 M_{1^6}\big)
\nonumber
\\
&\qquad
- 35(2M_{3,1}+ M_{1^4} )(M_{3,1} - M_{1^4})
\Big\}
\\
& -\frac{8}{2025}\big(45 s_5s_1-45 s_3^2-15 s_3s_1^3+s_1^6)t_1^2
    \nonumber
    \\
    & \times  \Big\{ \big(9 M_{5,1^3} - 5M_{3^2,1^2} - 5 M_{3,1^5}  + M_{1^8}  
    \big)
    \nonumber
    \\
&\qquad - M_{1^2} (9 M_{5,1} + 5M_{3^2} - 10 M_{3,1^3}  -4 M_{1^6}      \big)
    \nonumber
    \\
    &   \qquad + 5(2M_{3,1} + M_{1^4})(M_{3,1} - M_{1^4})
    \Big\}
    \label{BKP-even-c}
\\
& + \text{higher order}.
\nonumber
\end{align}
\end{subequations}
This equation holds identically in $\bs{s},\bs{t}$ and leads to
the quadratic relation 
$9 M_{5,1} - 5M_{3,3}  - 5 M_{3,1^3}  + M_{1^6} 
+  15 M_{1^2}(M_{3,1}-M_{1^4})$ at order 6
and two quadratic relations at order 8,
and so on.

The rhs of \eqref{Zt-linear} has up to order 8 the expansion
\begin{subequations}
  \label{BKP-lin-B}
  \begin{align}
0&=
-\frac{8}{2025}\big(
45 t_5t_1-45 t_3^2 -15t_3t_1^3
+t_1^6
\big)
\big(9M_{5,1} -5 M_{3^2}-5 M_{3,1^3} + M_{1^6}\big)
\\
&-\frac{4}{99225} \big(630 t_1t_7-630 t_3t_5
-105 t_1^3t_5  -315 t_1^2t_3^2+21t_1^5t_3 + t_1^8 
\big)
\nonumber
\\*[-1ex]
&\qquad\qquad
\times \big(90 M_{7,1}-42 M_{5,3}-21 M_{5,1^3}-35 M_{3^2,1^2}+7M_{3,1^5}+ M_{1^8}
\big)
\\
& + \text{higher order}.
\nonumber
\end{align}
\end{subequations}
The relation holds identically in $\bs{t}$ so that 
$0=\big(M_{1^6}-5 M_{1^3,3} -5 M_{3^2}-9M_{1,5}\big)$ at order 6,
and so on.

We also list the identities  \eqref{int-even} up to order $8$.
We rescale by $3/2$ at $k+m=4$, by
$45/2$ at $k+m=6$ and by $315/2$
at $k+m=8$ to achieve integer coefficients. The lhs below corresponds
to the rescaled lhs of  \eqref{int-even}. The rhs starts with $m=1$;
every next line has $m$ increased by $2$:
\begin{subequations}
\label{BKP-lin-C}
  \begin{align}
  2M_{3,1}+M_{1^4} &= M_{3,1}+ 2M_{1^4}
  \nonumber\\
  &= 3M_{3,1},
  \\
18 M_{5,1}+5 M_{3^2}&+20 M_{3,1^3}+2 M_{1^6}
\nonumber
\\
&=9 M_{5,1}+30 M_{3,1^3}+6 M_{1^6}
\nonumber
\\
&=15 M_{3^2}+30 M_{3,1^3}
\nonumber
\\
&= 45 M_{5,1},
\\
90 M_{7,1} + 42 M_{5,3} &+84 M_{5,1^3} + 70  M_{3^2,1^2} 
+28 M_{3,1^5}+ M_{1^8}
\nonumber
\\
&=45 M_{7,1} + 126 M_{5,1^3} + 70  M_{3^2,1^2} 
+70 M_{3,1^5} +4 M_{1^8}
\nonumber
\\
&=63 M_{5,3} + 210  M_{3^2,1^2} 
+42 M_{3,1^5}
\nonumber
\\
&=105 M_{5,3} + 210  M_{5,1^3} 
\nonumber
\\
&=315 M_{7,1}.
\end{align}
\end{subequations}
The relations (\ref{BKP-lin-C}) combine at order 4 and 6 to
\begin{align}
\label{lin-O46}
M_{1^4} &=M_{3,1}, 
\nonumber
\\
M_{1^6} &= - \tfrac{3}{2} M_{5,1} +\tfrac{5}{2} M_{3^2},
&
M_{3,1^3} &= \tfrac{3}{2} M_{5,1}- \tfrac{1}{2} M_{3^2}.
\end{align}
These are well-known \cite[eq.\ (49)]{Mironov:2020tjf} for the trivial
even potential
$V_0(x)=0$. That they hold for any
$V_0(x)=V_0(-x)$, and that they are part of infinite families of relations,
seem to be new results.
For $V_0(x)=\frac{1}{4}x^4$, explicit formulae for these moments can be found
in $1/N$-expansion along the lines of \cite{Branahl:2020yru}.

One checks that the linear relations (\ref{BKP-lin-B}) and
(\ref{BKP-lin-C}) alone do \emph{not} imply
the quadratic BKP equations (\ref{BKP-even}. The
linear relations reduce
(\ref{BKP-even-c})
to one further \emph{linear} identity
\begin{align}
0 &= 9 M_{5,1^3} - 5M_{3^2,1^2} - 5 M_{3,1^5}  + M_{1^8}  \;,
\label{BKP-even-o8-new}
\end{align}
which cannot be deduced from  (\ref{BKP-lin-B}) and
(\ref{BKP-lin-C}). We have verified that (\ref{BKP-even-o8-new})
together with the extension of (\ref{BKP-lin-B}) and
(\ref{BKP-lin-C}) to order 10 reduce the quadratic relation
(\ref{BKP-ZN-q})
for $V_0(x)=V_0(-x)$ at order 10 to linear relations
(\ref{BKP-even-new}) of order $n=10$.
It would be interesting to investigate whether this is always possible:
\begin{question}
For even potential $V_0(x)=V_0(-x)$, is the following true?

The quadratic  BKP relations \eqref{BKP-ZN-q} together with
the linear relations \eqref{Zt-linear} and \eqref{int-even} are
equivalent to
\eqref{Zt-linear} and \eqref{int-even} together with
further linear relations of order $n$ (even)
\begin{align}
0 &=\sum_{k=1}^{n-2r}(-1)^{k}
q_{k}(\bs{s})
\int_{\mathcal{H}_N} \!\!\! \dd\mu_{\Lambda,\bs{0}}(H)\,
q_k(\tilde{\bs{H}})
q_{n-2r-k}(\tfrac{1}{2}\bs{sH})
q_{2r}(\tfrac{1}{2}\bs{tH}),
\label{BKP-even-new}
\end{align}
for $r=1,...,\frac{n-6}{2}$ and identically in $\bs{t},\bs{s}$.
\end{question}

The relations (\ref{BKP-even-new}) for $n\mapsto n'$ are the restriction of the
BKP-relation \eqref{BKP-ZN-q} of order $n'$ to the linear part
$l=n=s=0$ and $r\geq 2$ (because $r=0$ is 
captured by \eqref{Zt-linear}; contributions to \eqref{BKP-ZN-q}
are of even order $\geq 6$ in $\bs{s}$). The relation at order $n'$
holds if recursively there are sufficiently many relations of order
$\leq n'-2$ which let the
mixed cases $l+n+s \neq 0$ and $k+m+r \neq 0$ in \eqref{BKP-ZN-q}
vanish identically.

\section*{Acknowledgements}

I am grateful to Ga\"etan Borot for the joint collaboration that
established the BKP relations for the Kontsevich matrix model with
arbitrary potential. I would like to thank Adam Afandi for his
SageMath programme which led me to the conjecture that linear relations
could hold. These relations are currently studied together with
Katharina Harengel; I thank her for the discussions.
This work was funded by the Deutsche
  Forschungsgemeinschaft (DFG, German Research Foundation) --
  Project-ID 427320536 -- SFB 1442, as well as under Germany's
  Excellence Strategy EXC 2044 390685587, Mathematics M\"unster:
  Dynamics -- Geometry -- Structure.


\begin{thebibliography}{DJKM82}
\expandafter\ifx\csname url\endcsname\relax
  \def\url#1{\texttt{#1}}\fi
\expandafter\ifx\csname doi\endcsname\relax
  \def\doi#1{\burlalt{doi:#1}{http://dx.doi.org/#1}}\fi
\expandafter\ifx\csname urlprefix\endcsname\relax\def\urlprefix{URL }\fi
\expandafter\ifx\csname href\endcsname\relax
  \def\href#1#2{#2}\fi
\expandafter\ifx\csname burlalt\endcsname\relax
  \def\burlalt#1#2{\href{#2}{#1}}\fi

\bibitem[BHW22]{Branahl:2020yru}
J.~Branahl, A.~Hock, and R.~Wulkenhaar.
\newblock {Blobbed topological recursion of the quartic Kontsevich model I:
  Loop equations and conjectures}.
\newblock {\em Commun. Math. Phys.}, 393(3):1529--1582, 2022,
  \burlalt{2008.12201}{http://arxiv.org/abs/2008.12201}.
\newblock \doi{10.1007/s00220-022-04392-z}.

\bibitem[BW24]{Borot:2023thu}
G.~Borot and R.~Wulkenhaar.
\newblock {A note on BKP for the Kontsevich matrix model with arbitrary
  potential}.
\newblock {\em SIGMA}, 20:050, 2024,
  \burlalt{2306.01501}{http://arxiv.org/abs/2306.01501}.
\newblock \doi{10.3842/SIGMA.2024.050}.

\bibitem[DJKM82]{Date:1981qz}
E.~Date, M.~Jimbo, M.~Kashiwara, and T.~Miwa.
\newblock {Transformation groups for soliton equations. 4. A new hierarchy of
  soliton equations of KP type}.
\newblock {\em Physica D}, 4:343--365, 1982.
\newblock \doi{10.1016/0167-2789(82)90041-0}.

\bibitem[HB21]{harnad_balogh_2021}
J.~Harnad and F.~Balogh.
\newblock {\em Tau Functions and their Applications}.
\newblock Cambridge Monographs on Mathematical Physics. Cambridge University
  Press, 2021.
\newblock \doi{10.1017/9781108610902}.

\bibitem[MM21]{Mironov:2020tjf}
A.~Mironov and A.~Morozov.
\newblock {Superintegrability of Kontsevich matrix model}.
\newblock {\em Eur. Phys. J. C}, 81(3):270, 2021,
  \burlalt{2011.12917}{http://arxiv.org/abs/2011.12917}.
\newblock \doi{10.1140/epjc/s10052-021-09030-x}.

\end{thebibliography}

\end{document}